\documentclass[conference]{IEEEtran}
\IEEEoverridecommandlockouts
\usepackage{cite}
\usepackage{amsmath,amssymb,amsfonts}
\usepackage{algorithmic}
\usepackage{graphicx}
\usepackage{textcomp}
\usepackage{xcolor}
\def\BibTeX{{\rm B\kern-.05em{\sc i\kern-.025em b}\kern-.08em
    T\kern-.1667em\lower.7ex\hbox{E}\kern-.125emX}}
\usepackage[T1]{fontenc}
\usepackage{algorithm}
\usepackage{algorithmic}
\usepackage{amssymb}
\usepackage{hyperref}
\usepackage{amsmath}

\usepackage{cite}

\usepackage{color}

\newtheorem{theorem}{Theorem}
\newtheorem{lemma}{Lemma}
\newtheorem{remark}{Remark}
\begin{document}

\title{Location-Restricted Stable Matching}

\author{\IEEEauthorblockN{Garret Castro}
\IEEEauthorblockA{\textit{University of California, Merced} \\
\textit{Merced, USA} \\
gcastro35@ucmerced.edu}
}

\maketitle

\begin{abstract}
Motivated by group-project distribution, we introduce and study stable matching under the constraint of applicants needing to share a location to be matched with the same institute, which we call the Location-Restricted Stable Matching problem (LRSM). We show that finding a feasible matching is NP-hard, which implies that finding a feasible and stable matching is also NP-hard. We then analyze the subproblem where all institutes have the same capacity $c$, and the applicant population of each location is a multiple of \textit{c}, which mimics more realistic constraints and makes finding a feasible matching in P. Even under these conditions, a stable matching (a matching without blocking pairs) may not exist, so we look for a matching that minimizes the number of blocking pairs. We find that the blocking pair minimization problem for this subproblem is inapproximable within n\textasciicircum(1-epsilon) for \textit{n} agents and extend this result to show that the problem of minimizing the number of agents in blocking pairs is also inapproximable within n\textasciicircum(1-epsilon). To address this complexity, we analyze the case where agents can only form blocking pairs with agents ``in'' their location, which, given a feasible matching, can be solved in O(n\textasciicircum2).
\end{abstract}

\begin{IEEEkeywords}
stable matching problems,
almost stable matching,
approximation algorithms,
combinatorics and graph theory,
combinatorial optimization,
complexity theory
\end{IEEEkeywords}

\section{Introduction}
In the admissions problem introduced by Gale and Shapley \cite{gale_college_1962}, students have a total order of colleges that they prefer to attend, and colleges have a total order of students they'd like to admit, as well as a capacity of how many students they can admit. The objective of the problem is to create a \textit{stable matching}, a perfect matching between students and colleges such that no student and college pair prefer each other to their matches. If a matching has such a student and college that prefer each other to their match, the student and college want to deviate from their matching, which ``blocks'' the matching's stability, so such a student-college pair is thus called a \textit{blocking pair}.

Gale and Shapley \cite{gale_college_1962} proved that a stable matching always exists, and one can always be found in $O(n^2)$ using what's now known as the \textit{Gale-Shapley Algorithm} (GS). However, certain matching restrictions from real-world constraints make finding a stable matching NP-hard, such as when colleges have upper and lower quotas instead of a capacity \cite{hamada_hospitalsresidents_2016}, when the total orderings can be incomplete and have ties (specifically for finding maximum-size stable matchings) \cite{wiedermann_stable_1999}, or when students can apply as couples \cite{ronn_np-complete_1990}.

One unexplored real-world restriction is the problem of compatibility between applicants. For example, engineering students at the University of California, Merced, take a mandatory capstone project course. Students have different preferences over which project they're assigned to, and clients hosting the project prefer that the assigned students have certain project development skills. However, there is an additional condition that each student is a member of a lab, and students who don't share a lab cannot work on the same project. Then, there is a restriction on which students can be matched to a project, based on the lab compatibility with other assigned students. This compatibility notion can be extended to other instances of mutual incompatibility in matching, such as network devices needing to share a protocol to communicate.

In this paper, we call an instance of the many-to-one stable matching problem with mutual applicant compatibility restrictions the \textit{Location-Restricted Stable Matching Problem} (LRSM for short), where applicants matched with an institute (or in our instance, students assigned to a project) must share a location with the another applicants matched to that institute. First, we show that finding a feasible matching (a perfect matching that satisfies lab restrictions, more precisely defined in Section 2) for LRSM is NP-hard. 

\begin{remark}
    For clarification, though the terminology is similar, this is not to be confused with the \textit{locally stable matching} of \cite{arcaute2009social}, where social connections affect matching stability but not feasibility.
\end{remark}

We then analyze LRSM instances assuming feasible matchings are easy to find. We show in Section 2 that the existence of a feasible matching does not imply the existence of a feasible and stable matching. However, it is typical in project-based courses that every student must be assigned a project. We are thus motivated to find the feasible matching with the least blocking pairs possible, or as coined by \cite{biro_almost_2012}, an ``almost stable'' matching. Developing algorithms to find such matchings is popular in the stable matching literature \cite{abraham2005almost, biro_size_2010, biro_almost_2012, hamada_hospitalsresidents_2016, khuller1994line}. We then analyze LRSM under the restriction that projects have the same capacity, and the location populations are evenly divisible by that capacity, restrictions that make a feasible matching solvable in polynomial time. We prove that even under these conditions, finding a stable matching is NP-hard within a factor of $n^{1-\epsilon}$ blocking pairs compared to the fewest number of blocking pairs, where $n$ is the number of agents (students and projects). We can extend this result to the problem of minimizing the number of \textit{blocking agents}, which is the number of agents involved in blocking pairs.

We propose a way to tackle this hardness by ignoring blocking pairs between a student and a project if the project is matched with students from a different location. Such blocking pairs in practical scenarios are likely less detrimental to the overall matching, as deviation is usually infeasible due to the location restrictions. We call stability under this notion \textit{l-stability}. Unlike a classical stable matching, we show that the existence of a feasible matching of LRSM implies the existence of an l-stable matching, and we present a polynomial-time algorithm to find such a matching.

\section{Preliminaries}
\subsection{Basic Definitions}

An instance of \textsc{LRSM} $I$ consists of a set of students that each belong to a location in the set of locations, a set of projects, each with capacities such that the sum of the capacities of projects in $I$ is equal to the number of students in $I$. Each student and project in $I$ has a strict total ordering for the other agent type (students prefer projects, and vice versa). 

We now present definitions of matchings in LRSM. We define a \textit{feasible matching} of LRSM as a many-to-one matching between students and projects such that each project is matched with exactly the number of students of its capacity, each student is matched with one project, and all students matched to a project are collocated (i.e., have the same location). We define a \textit{stable matching} of LRSM as a feasible matching with no blocking pairs, i.e., no student $s(\in S)$ prefers a project $p$ to its match such that $p$ also prefers $s$ to one of its matched students. Finally, we define an \textit{l-stable} matching as a feasible matching without blocking pairs between collocated students.

We define the problem of finding the feasible matching with the minimum number of blocking pairs \textit{Min-BP LRSM}. We define the problem of finding the feasible matching with the minimum number of \textit{blocking agents}, the number of agents involved in blocking pairs, \textit{Min-BA LRSM}. We say an algorithm $A$ is an $r(n)$-approximation for an LRSM problem if for the number of blocking pairs it generates in the worst-case, $A(x)$, and the number of blocking pairs in the optimal solution to Min-BP LRSM, $opt(x)$, $A(x)/opt(x) \leq r(n)$ for any instance $x$ of size $n$.

\subsection{Hard Restrictions on LRSM Problems}
\label{sec:restrictions}

As will be proven in Section \ref{sec:feasibility}, finding a feasible matching for LRSM is NP-hard, so immediately finding a stable matching is NP-hard. However, even in scenarios where feasibility is easy, such as those described in the following, finding a stable matching is still NP-hard. In fact, the hardness of Min-BP LRSM and Min-BA LRSM still hold given the following restrictions:

\begin{description}
    \item[(A1)] \textbf{Universal Project Capacity} Each project can be matched with the same number of students.
    \item[(A2)] \textbf{Evenly Divisible Local Populations} Each location has a number of locals that is a positive integer multiple of the universal project capacity.
    \item[(B1)] \textbf{Smallest Local Populations} Each location has only 2 locals.
    \item[(B2)] \textbf{Location Master List} All collocated students have the same preference list (i.e., are based on a ``master list'').
    \item[(B3)] \textbf{Project Master List} All projects have the same preference list.
\end{description}

An instance of Min-BP LRSM with restrictions A1 and A2 will be referred to as \textit{Min-BP Divisible LRSM}. Instances of Min-BP Divisible LRSM arise commonly in practical scenarios, such as when a teacher needs to distribute a fixed number of projects to evenly-sized teams of students. Note that finding a feasible matching in Min-BP Divisible LRSM is in P; since location populations are divisible by the universal project capacity $c$, students from the same location can be arbitrarily grouped into $c$-sized groups, then the groups can be matched with projects using a maximum one-to-one bipartite matching, which is in P. We call an instance of Min-BP LRSM with restrictions B1, B2, and B3 \textit{L2 Min-BP Divisible ML-LRSM} (with L2 indicating each location has population 2, and ML representing that student and project preferences come from master lists). We define \textit{L2 Min-BA Divisible ML-LRSM} analogously.

\subsection{A Starting Example}
An instance of LRSM with feasible matchings but no stable matching is shown in Figure \ref{fig:no-stable-matching}, hence the need for a minimization algorithm. In Figure \ref{fig:no-stable-matching}'s instance of LRSM, in a feasible matching, students $s_1$ and $s_2$ must be assigned the same project, and the same goes for $s_3$ and $s_4$. If $s_1$ and $s_2$ are assigned to project $p_1$, ($s_2$, $p_2$) will be a blocking pair. If assigned to $p_2$, then ($s_1$, $p_1$) will be a blocking pair.

\section{Feasibility}
\label{sec:feasibility}

\begin{theorem}
    Finding a feasible matching for LRSM is NP-complete. 
\end{theorem}
\label{feasibility-theorem}

\renewcommand{\arraystretch}{1.2}
\begin{figure}[h]
    \centering
    \begin{tabular}{c c c | c c c}
        \hline
        S. & Pref. & L. & P. & Pref. & C. \\
        \hline
        $s_1$ : & $p_1 \; p_2$ & A & $p_1$ : & $s_1 \; s_2...$ arbitrary & 2 \\
        $s_2$ : & $p_2 \; p_1$ & A & $p_2$ : & $s_2 \; s_1...$ arbitrary & 2 \\
        $s_3$ : & arbitrary & B &  &  &  \\
        $s_4$ : & arbitrary & B &  &  &  \\
        \hline
    \end{tabular}
    \caption{An instance of LRSM where no feasible and stable matching exists. The table indicates each student's preferences and location (S., Pref., and L., respectively), as well as each project's preferences and capacity (P., Pref., and C., respectively).}
    \label{fig:no-stable-matching}
\end{figure}

Membership in NP is obvious. We can prove NP-hardness by a reduction from a variation of the NP-hard problem \textsc{3-Partition} \cite{garey2002computers}. In an instance of 3-Partition, we are given an integer $m$, a multiset of integers $A=\{a_1, a_2, ... a_{3m}\}$ and an integer $T$ such that $\sum_{j=1}^{3m} a_j = mT$. For the instance we must find a set of $m$ disjoint triplets of elements of $a$ such that the sum of elements in each triplet is $T$. This remains NP-hard even if $\frac{T}{4} < a_j < \frac{T}{2} \quad \forall j$. 

Let $I_0=(m_0,A_0,T_0)$ be an instance of 3-Partition where $\frac{T_0}{4} < a_j < \frac{T_0}{2} \quad \forall j$. We create an instance $I$ of LRSM where for each element $a_j$ in $A_0$, we create a project with capacity $a_j$. We create $m_0$ locations, and $T$ students in each location. This reduction is obviously in polynomial time.

If there is a 3-partition for $I_0$, we can find a feasible matching for $I$. For each triplet of items in the solution to $I_0$, we arbitrarily pick a location and arbitrarily match its $T$ collocated students to the projects corresponding to the items in the triplet. As the total capacities of the items in triplets are $T_0$ (by definition of a 3-partition), this matching does not violate location or capacity restrictions. Since there are $m_0$ locations whose locals are distributed to 3 of the $3m$ projects, each project and student is matched, thus creating a feasible matching.

If there is a feasible matching $M$ for $I$, we can find a 3-partition for $I_0$. Slightly abusing vocabulary, we say a project is matched with a location $l$ if its students are all in $l$. We create a set for each location $l$ containing the projects matched with $l$. We know there are $m_0$ such sets because there are $m_0$ locations. Since in a feasible matching every student is matched with one project, each project is matched with the same number of students as its capacity, and there are $T_0$ students at each location, the sum of the capacities of the projects in each set is $T_0$. Because $\frac{T_0}{4} < a_j < \frac{T_0}{2} \quad \forall j$, it is easy to see that exactly 3 projects are matched with each location, and thus that each set has 3 projects. By constructing sets based on the corresponding item from $A$, it is easy to see that each new set sums to $T_0$. Since each new set has 3 items and there are $m$ sets total, this is a 3-partition for $I_0$.
\hfill \IEEEQED

\section{Hardness of Divisible LRSM Problems}

\subsection{Inapproximability of Min-BP Divisible LRSM}
\begin{theorem}
\label{approx-limit-theorem}
For any $\epsilon>0$, there is no polynomial-time $n^{1-\epsilon}$-approximation algorithm, where $n$ is the number of agents, for an instance of L2 Min-BP Divisible ML-LRSM, unless P=NP.
\end{theorem}

\emph{Proof}  This proof follows the structure of \cite{hamada_hospitalsresidents_2016}'s Theorem 1, which proves the inapproximability of minimizing the number of blocking pairs when projects have lower and upper quotas. They prove the approximation hardness using a polynomial-time reduction from the NP-complete problem \textit{Vertex Cover} (VC) \cite{garey2002computers}. While maintaining the core strategy, we modify the approach to fit our constraints. 

Since each student in a location will have the same preference, we will refer to the shared preference list as the preference list of the location. We refer to the two students at each location as the ``local pair'' $s$, or individually as $s^+$ and $s^-$, or the positive and negative members of $s$, respectively. We refer to the sets containing only the positive or negative members of each $s \in S$ as $S^+$ and $S^-$, respectively. \begin{remark}
    Most stable matching papers usually refer to elements such as $s$ as a single agent. However, due to the unique constraint in our reduction's instance that local pairs of agents that cannot be matched to different projects, for this proof only, we use $s$ to refer to a local pair of agents for conciseness. If, for this proof, we ever need to refer to a single student, we refer to it as $s^+$ or $s^-$, or, depending on the relevance, simply ``a student in $s$.''
\end{remark}

Given a VC instance $I_0 = (G_0, K_0)$, where $G_0 = (V_0, E_0)$ and $K_0$ is a positive integer, the goal is to find a subset of vertices $V_{0c}$ that ``covers'' each edge, i.e. each edge is connected with at least one vertex in $V_{0c}$, such that $|V_{0c}|=K_0$. We reduce $I_0$ to an instance of L2 Min-BP Divisible ML-LRSM, $I$. Let $n_v=|V_0|$, $c=\lceil{\frac{8}{\epsilon}}\rceil$, $B_1={n_v}^c$, and $B_2=\frac{1}{2}{n_v}^c-|E_0|$. Let the set of all students in $I$ be $S = C \cup F \cup G \cup X$ and the set of all projects be $P = V \cup H \cup Y$, with each subset defined in Figure \ref{fig:agent-defs}, with each element of $S$ representing a local pair. 

Let $[A_0]$ be a total ordering of agents for any set of agents $A_0\subset A$. The preferences of the students are defined in Figure \ref{fig:agent-prefs}, and the preference list of each project is $[X^+]\;[C]\;[G*]\;[F]\; [X^-]$. $[G*]$ is a total ordering of $[G^{i,j}*]$ for each $G^{i,j} \in G$ in any order, where $[G^{i,j}]$ is an order of students $g^{i,j}_{b,a}$ sorted by $b$ then $a$ ascending, with the exception of $g_{1,1}^{i,j}$, which at the beginning of $[G^{i,j}*]$.

Slightly abusing vocabulary, we say that a local pair $s$ is ``matched'' with a project $p$ if the students in $s$ are matched with $p$. By the location restriction, if in a feasible matching $M$ one student in $s$ is matched with $p$, both students are matched to $p$ in $M$. We also say $s$ is ``in'' the set, both its member students are in. Note that if one member of $s$ is in a set, both members are in the set.

\begin{figure}[h]
    \centering
    \rule[0mm]{\columnwidth}{0.3mm}
    
    \scalebox{0.75}{
    $\displaystyle
        \begin{array}{lll}
            S &= C \cup F \cup G \cup X \\
            C &= \{ c_i \mid 1 \leq i \leq K_0 \} & \\
            F &= \{ f_i \mid 1 \leq i \leq {n_v} - K_0 \} & \\
            G^{i,j} &= \left\{ g^{i,j}_{0,a} \mid 1 \leq a \leq B_2 \right\} \cup \left\{ g^{i,j}_{1,a} \mid 1 \leq a \leq B_2 \right\} & (v_i, v_j) \in E_0, i < j \\
            G &= \bigcup G^{i,j} \\
            X &= \{ x_i \mid 1 \leq i \leq B_1 \} &  \\
            \\
            P &= V \cup H \cup Y \\
            V &= \{ v_i \mid 1 \leq i \leq {n_v} \} & \\
            H^{i,j} &= \left\{ h^{i,j}_{0,a} \mid 1 \leq a \leq B_2 \right\} \cup \left\{ h^{i,j}_{1,a} \mid 1 \leq a \leq B_2 \right\} & (v_i, v_j) \in E_0, i < j \\
            H &= \bigcup H^{i,j} \\
            Y &= \{ y_i \mid 1 \leq i \leq B_1 \} \\
        \end{array}
    $
    }
    
    \rule[0mm]{\columnwidth}{0.3mm}
    \caption{Definitions for each set of agents in $I$. Note that in these definitions, each element in $S$ is a local pair.}
    \label{fig:agent-defs}
\end{figure}

\begin{figure}[h]
    \centering
    \rule[0mm]{\columnwidth}{0.3mm}
    
    \scalebox{0.75}{
    $\displaystyle
        \begin{array}{lllll}
            c_i & : \quad [[V]] && [[Y]] \quad \dots &  (1 \leq i \leq K_0) \\
            f_i & : \quad [[V]] && [[Y]] \quad \dots & (1 \leq i \leq {n_v} - K_0) \\
            g_{0,1}^{i,j} & : \quad h_{0,1}^{i,j} & v_i \quad h_{1,1}^{i,j} & [[Y]] \quad \dots & ((v_i, v_j) \in E_0, i < j) \\
            g_{0,2}^{i,j} & : \quad h_{0,2}^{i,j} & v_i \quad h_{0,3}^{i,j} & [[Y]] \quad \dots & ((v_i, v_j) \in E_0, i < j) \\
            & \quad \vdots  \\
            g_{0,B_2-1}^{i,j} & : \quad h_{0,B_2-1}^{i,j} & v_i \quad h_{1,B_2}^{i,j} & [[Y]] \quad \dots & ((v_i, v_j) \in E_0, i < j) \\
            g_{0,B_2}^{i,j} & : \quad h_{0,B_2}^{i,j} & v_i \quad h_{0,1}^{i,j} & [[Y]] \quad \dots & ((v_i, v_j) \in E_0, i < j) \\
            g_{1,1}^{i,j} & : \quad h_{0,2}^{i,j} & v_j \quad h_{1,2}^{i,j} & [[Y]] \quad \dots & ((v_i, v_j) \in E_0, i < j) \\
            g_{1,2}^{i,j} & : \quad h_{1,2}^{i,j} & v_j \quad h_{1,3}^{i,j} & [[Y]] \quad \dots & ((v_i, v_j) \in E_0, i < j) \\
            & \quad \vdots \\
            g_{1,B_2-1}^{i,j} & : \quad h_{1,B_2-1}^{i,j} & v_j \quad h_{1,B_2}^{i,j} & [[Y]] \quad \dots & ((v_i, v_j) \in E_0, i < j) \\
            g_{1,B_2}^{i,j} & : \quad h_{1,B_2}^{i,j} & v_j \quad h_{1,1}^{i,j} & [[Y]] \quad \dots & ((v_i, v_j) \in E_0, i < j) \\
            x_i & : \quad y_i && [[Y\;\backslash\;y_i]] \quad \dots & (1 \leq i \leq B_1)  \\
        \end{array}
    $
    }
    
    \rule[0mm]{\columnwidth}{0.3mm}
    \caption{The preferences of local pairs in $I$}
    \label{fig:agent-prefs}
\end{figure}

Note that $|P|=2|S|$, otherwise it will be impossible to perfectly match students to projects, as projects have capacity 2 (this can also be checked arithmetically from the sizes given in Figure \ref{fig:agent-defs}). As $|P|={n_v}+2|E_0|B_2+B_1$, the number of agents $|P|+|S|=n$ is $n=3({n_v}+2|E_0|B_2+B_1)=3((|E_0|+1){n_v}^c+{n_v}-2|E_0|^2) < 3(2{n_v}^{c+2}+{n_v}^c+{n_v}) \leq 3(4{n_v}^{c+2}) = 12{n_v}^{c+2}$, which is polynomial ${n_v}$.

\subsubsection{Preparatory Lemmas}
In the following, we use $[Y]$ to refer to both $[Y]$ and $y_i \cup Y\;\backslash\;y_i $ for convenience.

\begin{lemma}
\label{lemma:y_x_pair}
A matching between a local pair not in $X$ and a project in $Y$ has at least $B_1$ blocking pairs.
\end{lemma}
\textit{Proof} Suppose there is a matching where a local pair $s \not\in X$ is matched with a project in $Y$. Since each project is matched with exactly one local pair, and the number of local pairs in $X$ is equal to $|Y|$, it follows that at least one local pair in $X$, say $x$, is not assigned to a project in $Y$. $x^+$, then, will form a blocking pair with every project in $Y$; and $|Y|=B_1$. \hfill \IEEEQED
\begin{lemma} \label{lemma:prohibited_pair}
A matching where a student is assigned to a project to the right of $[Y]$ in their preference list has at least $B_1$ blocking pairs.
\end{lemma}
\textit{Proof} Suppose for some matching $M$ a local pair $s$ is matched with a project $p$ that is to the right of $[Y]$ in their preference list, and that there are less than $B_1$ blocking pairs. 

First note that if in $M$ a student not in $X$ is matched to a project in $Y$ there are $B_1$ blocking pairs via Lemma \ref{lemma:y_x_pair}. This means that in $M$ only local pairs in $X$ are matched with $Y$.

If $s$ is matched to a project to the right of $[Y]$ in its preference list, its members will form blocking pairs with each project in $Y$; each project in $Y$ prefers $s^+$ and $s^-$ to its matched member of $X^-$. This forms $2|Y|=2B_1>B_1$ blocking pairs, a contradiction. Therefore, a matching cannot both have a student assigned to the right of [Y] in their preference list and have less than $B_1$ blocking pairs. \hfill \IEEEQED

Based on Lemma \ref{lemma:y_x_pair} and Lemma \ref{lemma:prohibited_pair}, we will refer to any matching between a local pair $s$ and a project to the right of $[Y]$ in $s$'s preference list, or any matching between a local pair $s(\notin X)$ and a project in $Y$, as a ``prohibited pair.''

We will refer to each pair of student set and project set $G^{i,j}$ and $H^{i,j}$ an edge gadget, $g_{i,j}$. A matching ``of'' $g_{i,j}$ is a perfect matching between local pairs in $G^{i,j}$ and projects in $H^{i,j}$, and a blocking pair ``in'' a matching of $g_{i,j}$ is a blocking pair ($s$, $p$) (where $s$ is a local pair) such that $s\in G^{i,j}$ and $p\in H^{i,j}$.

\begin{lemma} \label{lemma:edge-gadget-internal}
There are only two matchings of an edge gadget $g_{i,j}$ that don't contain prohibited pairs, and each has two blocking pairs in it.
\end{lemma}
\textit{Proof} For a more illustrative explanation of this result, readers may find the proof of Lemma 2 in \cite{hamada_hospitalsresidents_2016} illuminating.

First, there are two matchings within an edge gadget that don't have prohibited pairs. For each $1 \leq a \leq B_2$, we can either match local pair $G^{i,j}_{0,a}$ with its most preferred match, and $G^{i,j}_{1,a}$ with its third most preferred match, or the opposite, matching each $G^{i,j}_{0,a}$ with its third most preferred match and each $G^{i,j}_{0,a}$ with its most preferred match. 

Note that for the former, each $G^{i,j}_{1,a}$ prefers $v_j$ to its assignment and for the latter $G^{i,j}_{0,a}$ prefers to $v_i$ as its assignment. We will accordingly call the former the $v_j$-preferred matching of $g_{i,j}$ and the latter matching the $v_i$-preferred matching of $g_{i,j}$. 

By constructing a bipartite graph where each vertex set is $G^{i,j}$ and $H^{i,j}$ and the edges are between the $s(\in G^{i,j})$ and $p(\in H^{i^j})$ if and only if ($s$, $p$) is not a prohibited pair, we get a graph of a cycle length of $4|B_2|$. As there are $2B_2$ vertices in each set, there are only two perfect matchings, and they are the ones described above.

By inspection, the only blocking pairs in the $v_j$-preferred matching are ($g^{i,j+}_{1,1}$, $h^{i,j}_{0,2}$) and ($g^{i,j-}_{1,1}$, $h^{i,j}_{0,2}$), and the only blocking pairs of the $v_i$-preferred matching are ($g^{i,j+}_{0,1}$, $h^{i,j}_{0,1}$) and ($g^{i,j-}_{0,1}$, $h^{i,j}_{0,1}$).
\hfill\IEEEQED

\begin{lemma} \label{lemma:edge-gadget-external}
A matching $M$ with no prohibited pairs only has blocking pairs involving students in $G$ and projects not in $H$ if and only if one of the following is true of any edge gadget $g_{i,j}$:

\begin{enumerate}
    \item $g_{i,j}$'s matching is $v_i$-preferred and $v_i$'s matched local pair is not in $C$
    \item $g_{i,j}$'s matching is $v_j$-preferred and $v_j$'s matched local pair is not in $C$.  
\end{enumerate}
\noindent Moreover, if either is the case for any $g_{i,j}$, $M$ has at least $B_1$ blocking pairs.

\end{lemma}
\textit{Proof} We first note that if $M$ has no prohibited pairs, and $v_i$'s matched local pairs aren't in $C$, their matched local pairs must be in $F$. If $v_i$'s matching is in $F$, then it will prefer any student from $G$ to its matching, and local pairs $G^{i,j}_{0,a}$ (for $1\leq a \leq B_2$) prefer $v_i$ in the $v_i$-preferred matching, creating $2B_2$ blocking pairs. We know each edge gadget matching has two blocking pairs within it, and there are $|E_0|$ edge gadgets, creating a total of $2|E_0|$ blocking pairs within all the edge gadgets in $M$. Combined, there are $2B_2+2|E_0|=B_1$ blocking pairs in $M$.

Since any other project is to the right of the local pair $s$($\in G^{i,j}_{0,a}$)'s matching, there are no blocking pairs between $s$ and any project $p \in P\;\backslash\;(H\;\cup\;v_i)$. Additionally, if $v_i$ as matched with a local pair from $C$, $v_i$ will prefer its matching to all local pairs in $G$, so if $g_{i,j}$'s matching is the $v_i$-preferred one, there will be there will be no blocking pairs ($s$, $p$). The same reasoning can be used to prove the $v_j$-case with local pairs $G^{i,j}_{1,a}$. 
\hfill\IEEEQED

\subsubsection{Gap for Inapproximability}
\begin{lemma} \label{lemma:yes-vc}
If $I_0$ is a “yes” instance of VC, then $I$ has a solution with at most $2n^2+2|E_0|$ blocking pairs.
\end{lemma}
\textit{Proof} Suppose there is a vertex cover $V_{0c}$ such that $|V_{0c}|=|K_0|$. We can construct a matching for $M$ by first arbitrarily matching each local pair in $C$ with projects that correspond to vertices in $V_{0c}$, and each local pair in $F$ with the projects that correspond to the $V \backslash V_{0c}$. Since $|C \cup F|=2{n_v}$ and $|V_0|={n_v}$, there are at most $2{n_v}^2$ blocking pairs between $C\;\cup\;F$ and $V$. It is easy to see that students in $C\;\cup\;F$ prefer their matched projects to projects in $[Y]$ or to the right of $[Y]$ in their preference list, so they are involved in no other blocking pairs.

Then, we match $G$ and $H$ using perfect matchings within edge gadgets. By the definition of a vertex cover, for each edge $(v_i, v_j)$ where $(v_i, v_j) \in E_0, i < j$, one of the vertices is in the vertex cover and thus one of the corresponding projects is assigned a project in $C$. By Lemma \ref{lemma:edge-gadget-external}, if either $v_i$ or $v_j$ is in $C$, there is a matching that avoids blocking pairs between a student in $G^{i,j}$ and a project not in $H^{i,j}$. We select the matching that avoids such blocking pairs (the $v_i$-preferred if $v_i$ is assigned a local pair in $C$, otherwise the $v_j$-preferred matching). In these matchings, the only blocking pairs involving students in $G$ are within the $|E_0|$ edge gadgets, of which there are two each by Lemma \ref{lemma:edge-gadget-internal}. Thus, there are $2|E_0|$ blocking pairs involving students in $G$.

Lastly, we assign all local pairs in $X$ to the project at the top of their preference list. It is easy to see that, in this matching, neither local pairs in $X$ nor projects in $Y$ are involved in no blocking pairs.

Since students in $C\;\cup\;F$ are involved in at most $2{n_v}^2$ blocking pairs, students in $G$ are involved in exactly $2|E_0|$, and students in $X$ are involved in none, $M$ has at most $2{n_v}^2+2|E_0|$ blocking pairs in total.
\hfill \IEEEQED

\begin{lemma} \label{lemma:no-vc}
If $I_0$ is a “no” instance of VC, then $I$ has a solution with at least $B_1$ blocking pairs.
\end{lemma}

\textit{Proof} We show that if $I$'s solution $M$ has less than $B_1$ blocking pairs, $I_0$ has a vertex cover of size $K_0$. By Lemma \ref{lemma:y_x_pair} and Lemma \ref{lemma:prohibited_pair}, $M$ cannot have prohibited pairs. Thus, local pairs $C \cup F$ are matched one-to-one with projects in $V$, and local pairs in $X$ to $Y$. By Lemma \ref{lemma:edge-gadget-external}, if for an edge gadget $g_{i,j}$ neither $v_i$ nor $v_j$ is matched with a local pair in $C$, there are $B_1$ blocking pairs. Thus, either $v_i$ or $v_j$ are in $C$ for all $g_{i,j}$. Hence, each edge in $I_0$ is connected with a vertex that corresponds to a project matched with $C$. Definitionally, these vertices form a vertex cover, and its size is $|C|=|K_0|$.
\hfill\IEEEQED

Finally, we estimate the gap between the ``yes'' and ``no'' VC instances. As observed previously, $n\leq 12{n_v}^{c+2}$. Hence, $B_1/({n_v}^2+2|E_0|) \geq {n_v}^c/3{n_v}^2 = 27{n_v}^{c+2}3^{-4}{n_v}^{-4} \geq 27{n_v}^{c+2}{n_v}^{-8}>n^{1-\frac{8}{c}} \geq n^{1-\epsilon}$. Thus, a polynomial-time $n^{1-\epsilon}$-approximation algorithm for L2 Min-BP Divisible ML-LRSM solves VC, implying P=NP. This proves Theorem \ref{approx-limit-theorem}.
\hfill\IEEEQED

\subsection{Inapproximability and Approximability of Min-BA Divisible LRSM}

\begin{theorem}\label{approx-limit-theorem-agents}
    For any $\epsilon$ > 0, there is no polynomial-time $n^{1-\epsilon}$-approximation algorithm for instance of L2 Min-BA Divisible ML-LRSM, unless P=NP.
\end{theorem}
We can use a nearly identical construction as in the proof of Theorem \ref{approx-limit-theorem}, with the only difference being setting $c=\lceil{\frac{9}{\epsilon}}\rceil$. 

\begin{lemma} \label{yes-vc-ba}
    The matching used in Lemma \ref{lemma:yes-vc} creates at most $3{n_v}+6|E_0|$ blocking agents for the ``yes'' case.
\end{lemma}
All agents in $C\;\cup\;F\;\cup\;V$ can be a blocking agent, and $|C\;\cup\;F\;\cup\;V|=3{n_v}$. The 2 blocking pairs from Lemma \ref{lemma:edge-gadget-internal} have a combined 6 blocking agents, meaning there are $6|E_0|$ blocking agents in $G$ and $H$. Since there are no blocking pairs between students in $X$ and projects in $Y$, the total blocking agent count is $3{n_v}+6|E_0|$. \hfill\IEEEQED

\begin{lemma} \label{no-vc-ba}
    The matching restrictions found in Lemma \ref{lemma:no-vc} create at least $B_1$ blocking agents for the ``no'' case.
\end{lemma}
As seen in the proofs of Lemma \ref{lemma:y_x_pair} and Lemma \ref{lemma:prohibited_pair}, prohibited pairs create blocking pairs between all projects in $Y$, and thus there are $|Y|=B_1$ blocking agents if there is a prohibited pair. Additionally, the $2B_2$ blocking pairs seen between projects in $V$ and students in $G^{i,j}$ form at least $2B_2$ blocking agents, so combined with the $6|E_0|$ blocking agents in each $G^{i,j}$ from Lemma \ref{lemma:edge-gadget-internal}, there are $2B_2+6|E_0|>B_1$ blocking pairs.\hfill\IEEEQED

Estimating the gap between ``yes'' and ``no'' instances, we get $B_1/(3{n_v}+6|E_0|) \geq B_1/(9{n_v}^2)=27{n_v}^{c+2}3^{-5}{n_v}^{-4}\geq27{n_v}^{c+2}{n_v}^{-9}>n^{1-\frac{9}{c}} \geq n^{1-\epsilon}$. This proves Theorem \ref{approx-limit-theorem-agents}.
\hfill\IEEEQED

Because there are $n$ agents and therefore at worst $n$ potential blocking agents, the result of Theorem \ref{approx-limit-theorem-agents} is almost tight.

\subsection{Finding L-stable Matchings}
\begin{theorem}
\label{l-stable}
Given a feasible matching $M$ for an instance of LRSM $I$, there exists a polynomial-time algorithm to find an l-stable matching for $I$.
\end{theorem}
We provide the algorithm here, after introducing some vocabulary. Slightly abusing notation, we say for some matching $M$ that a project is “in” a location $l$ if its matched students belong to $l$ (leveraging the fact that in a feasible matching, all students matched to a project are from the same location). 

\begin{algorithm}
\caption{L-stable Matching from a Feasible Matching}
\begin{algorithmic}[1]
\label{alg}
\STATE{
    From the given feasible matching $M$, create $|L|$ (the number of locations) instances of the admissions problem by removing from each agent’s preference list all agents it is not collocated with in $M$.
}
\STATE{
    Use a polynomial-time stable matching algorithm (such as the Gale-Shapley algorithm [1]) on each admissions problem instance, then combine and output the resulting matchings.
}
\end{algorithmic}
\end{algorithm}

Step 1 is obviously in polynomial time.  Defining $n$ to be the number of agents, creating the admissions problem instances can be done in $n^2$, and performing $|L|$ stable matchings can be done in $O(n^2)$. Since $|L|$ is polynomial-time n, Step 2 can be done in polynomial time. Note that after the stable matching algorithm is performed, there are no blocking pairs containing collocated agents, meaning the matching is an l-stable one, thus completing the proof. \hfill \IEEEQED

\section{Concluding Remarks}

This paper covers an instance of LRSM that appears often in practical scenarios and makes the corresponding 3-Partition solvable in polynomial time. More investigation can be done into other practical variations of the 3-Partition problem that are also solvable in polynomial time.

There remains an open question as to whether or not the bound to Theorem 2 is tight. For example \cite{hamada_hospitalsresidents_2016}, the base for Theorem 2 and 3’s proofs, found a tight result by leveraging the Rural Hospitals Theorem, but there is currently no analogous theorem for our case. 

An interesting generalization is to let students hold memberships in several locations—akin to speaking multiple languages—so they can form groups if they share at least one location or language. There is an open question as to whether stronger approximation floors can be found for this case.

Location restrictions are not mutually exclusive with other stable matching restrictions, so it is worthwhile to find the minimum number of blocking pairs under restrictions combined with location restrictions, such as upper/lower quotas \cite{hamada_hospitalsresidents_2016}, ties \cite{wiedermann_stable_1999}, incomplete lists \cite{wiedermann_stable_1999}, and classifications \cite{huang_classified_2010}.

\cite{hamada_hospitalsresidents_2016} and \cite{biro_almost_2012} investigate the problem of minimizing the number of applicants (or students, in our case) involved in blocking pairs as opposed to the number of blocking agents, and further research can be done to find an approximation limit and algorithm for the number of blocking students for LRSM.

While much of this paper features negative complexity results, it establishes important theoretical foundations that enable future algorithmic developments. The tractable special cases of stable matching restrictions explored in \cite{biro_almost_2012} and the parameterized algorithms for blocking pair minimization developed in \cite{hamada_hospitalsresidents_2016} provide promising directions for extending these theoretical insights into practical algorithmic frameworks.

\section*{Acknowledgment}

I thank Santosh Chandrasekhar for introducing this problem and his invaluable mentorship.

\bibliographystyle{IEEEtran}
\bibliography{main, SMT-Diverse}

\end{document}